\documentclass[aps,twocolumn,noshowkeys,noshowpacs,nofootinbib]{revtex4-1}
\usepackage{epsf,amsmath,graphicx,amssymb,url,hyperref}
\begin{document}
\title{Weyl meson and its implications in collider physics and cosmology }
\author{Gopal Kashyap}
\email{gopal@iitk.ac.in}
\affiliation{Department of Physics, IIT Kanpur, Kanpur 208 016, India}

\begin{abstract}
Local scale invariant theory leads to the existence of a new particle
called the Weyl vector meson. We study a generalized Standard Model,
which displays local scale invariance. The model contains a real scalar
field, besides the Higgs multiplet and coupling between the Higgs and Weyl meson. 
For certain range of coupling parameters, 
 the Weyl-Higgs coupling leads to interesting
phenomenon in particle colliders as well as in cosmology. 
Here we study the signature of the Weyl meson in particle colliders and determine the range of coupling parameters for which it can solve the dark
matter problem. 
\end{abstract}
\maketitle
\section{Introduction}
One of the major unsolved problems in modern physics is the dark matter and the dark energy. 
The idea of dark matter came up with the discovery of inconsistency in 
galactic rotation curves and in analysis of galaxy clusters \cite{oort,zwiky,zwiky1,rubin}. Whereas, the dark energy is required in order to explain the accelerated expansion of the Universe\cite{riess,perlmutter}.
 Observations shows that $23\%$ of total energy density of the Universe is 
in the form of dark matter, $72\%$ is dark energy and rest is ordinary matter\cite{wmap1}. All of our scientific studies are related with
 this ordinary visible matter, which is a very small fraction of the Universe. 
The dark components of the Universe are still not well understood. 
There are many theoretical models dealing with dark energy, 
but here our main concern is the dark matter.
There currently exist many dark 
matter candidates \cite{bertone,jungman,bergstrom,feng,Bertone1}. One interesting possibility is the Weyl meson which arises naturally in local scale invariant theories. 
This symmetry was introduced by H. Weyl in 1929 \cite{weyl}, and the gauge particle associated with this symmetry is called the Weyl meson. Local scale 
invariance has been reviewed by many authors
\cite{dirac,sen,utiyama,utiyama1,Freund1,hayashi,hayashi1,
nishioka,ranganathan,rajpoot,rajpoot1} and the mechanism for breaking this symmetry has been discussed in Refs. \cite{jain1,pseudoscale,brokenscale}.
 In Refs. \cite{cheng,hcheng}, it has been shown that the Weyl meson acts
as a dark matter candidate in a locally scale invariant generalization
of the Standard Model.  
Cosmological implications of Weyl meson in this theory,
which contains only one Higgs doublet, 
 have been studied in \cite{jain1,DEDM,huang,wei}.
 But in this model Higgs particle get eliminated from the particle spectrum, which 
leads to violation of perturbative unitarity beyond electroweak scale, as shown in Refs. \cite{joglekar,cornwall}. 
Hence this model needs to be generalized. 

In Ref.\cite{quantumweyl} a generalization of this model has been proposed by
including an additional real scalar field $\chi$. In this generalized locally 
scale invariant Standard Model, the Higgs like particle 
remains in the physical spectrum and also couples directly with the Weyl meson. This model respects 
the unitarity condition and hence is
perturbatively stable. The phenomenological implications of this model are 
also very interesting and need to be studied in detail, which is the main motive of this work.

In this paper we started with the model described in Ref.\cite{quantumweyl}, which contains the coupling between the Weyl meson and
the Higgs boson. The Weyl meson, which interacts only with the Higgs particle and no other SM particle, may serve as a
good candidate for the dark matter. To strengthen this possibility we have to look for 
the processes, involving the Weyl meson, at the colliders and also in the Universe. The coupling strength, which plays a crucial 
role in all these processes, is determined by the free parameters of the theory, $f$ and $\beta_1$. 
We write the parameter $f$ in the term of Weyl meson's mass $M_S$ as, $M_S \approx fm_{pl}$ . Now our free parameters are $\beta_1$
and $M_S$.
 For the model introduced by Cheng \cite{cheng}, there is a cosmological constraint on the mass of the Weyl meson as shown in
Ref.\cite{huang}. Here we do not consider any unitarity bound on the 
mass of Weyl meson. Our main goal in this paper is to put some constraint on the value of coupling parameters $\beta_1$ and mass $M_S$ so 
that the Weyl meson can act as a suitable dark matter candidate.

This paper is organised as following. In Sec. II we start with  generalized local
scale invariant Standard Model and write down the interaction Lagrangian for Weyl meson. We discuss 
the implications of Weyl meson at colliders. For that we calculate the Higgs production cross section
at Large Hadron Collider (LHC) by including these interaction term. In Sec. III we consider the Weyl meson as dark matter
candidate and find the possible values of parameters for which it has relic density equal to 
dark matter energy density. In Sec. IV we further constrain these values by demanding the life time of Weyl meson 
to be equal or greater than the age of Universe. Finally we conclude in Sec. V.
 \section{Weyl meson and Higgs boson production at LHC: Implication of generalized local scale invariant 
Standard Model}
The action for the generalized locally scale invariant Standard Model, including an additional real scalar field 
$\chi$, may be written as \cite{quantumweyl,shapo}
\begin{equation}
\begin{split}
\mathcal{S}=\int& d^4x \sqrt{-\bar{g}} \left(\left[\frac{\beta }{8} \chi ^2+\frac{\beta _1}{4} 
{\mathcal{H}^\dagger} \mathcal{H} \right] {\bar{R}}'\right.\\
& \left.+{\bar{g}}^{\mu \nu} \left(D_\mu \mathcal{H} \right)^\dagger \left(D_\nu \mathcal{H} \right) +\frac{1}{2} {\bar{g}}^{\mu \nu}\left(D_\mu \chi 
\right) \left(D_\nu \chi \right) \right.\\ 
&\left. -\frac{1}{4}\lambda \chi ^4-\frac{1}{4}\lambda_1 \left[2{\mathcal{H}^\dagger}\mathcal{H}-{\lambda_2} {\chi}^2 \right]^2 \right)
\end{split}
\label{action}
\end{equation}
Here $\mathcal{H}$ is the Higgs doublet denoted as $\mathcal{H}=\mathcal{H}_0+\mathcal{\hat{H}} $, $ \mathcal{H}_0$ is the classical solution to 
the Higgs field, given as
\begin{align}
 \mathcal{H}_0=\frac{1}{\sqrt{2}}\left(
\begin{array}{c}
 0 \\
 v
\end{array}
\right) .
\end{align}
We can parameterize the $\mathcal{\hat{H}}$ in such a way that only its $ \phi_3$ component is non-zero.
The scale-covariant curvature scalar,${\bar{R}}'$, is defined in Ref.\cite{donoghue} and $D_\mu$ is the scale-covariant derivative
defined as
\begin{align}
 D_\mu=\partial_\mu-fS_\mu ,
\end{align}
where $f$ is the gauge coupling constant and $S_\mu $ is the Weyl meson field. The field $\chi$ in this model is a singlet under electroweak symmetry transformations. The potential shown in Eq.(\ref{action}) is stable due to underlying scale symmetry 
and also quantum corrections will not lead to any fine tuning problems, as shown in Ref.\cite{shapo}. By making the quantum expansion of the fields around its classical values, 
\begin{equation}
 \begin{split}
  \chi   & =\chi_{0}+\hat{\chi}, \; \; \;\;\;\;
\phi_{3} =\phi_{3,0}+\hat{\phi_{3}}\\
\bar{g}_{\mu\nu}&=g_{\mu\nu}+h'_{\mu\nu} \ ,
 \end{split}
\end{equation}
we can extract the quadratic mass and mixing terms of the fields.
The classical value of the field $\chi$ is taken of the order 
of Planck's mass and the graviton field is redefined, 
\begin{align}
 h'^\beta_\alpha={4 \over \sqrt{\beta \chi_0^2+\beta_1 v^2}} h^\beta_\alpha \ ,
\end{align}
so that its kinetic
energy term gets properly normalized.
Detailed quantum treatement of this model has been discussed
in Ref.\cite{quantumweyl}. 
There it was shown that after choosing a proper gauge one can eliminate the 
mixing terms and identify the 
mass term of Weyl meson as,
\begin{equation}
M_S^2=f^2\left[\chi_0^2\left(1+\frac{3\beta}{2}\right)+v^2\left(1+\frac{3\beta_1}{2}\right)\right].
\end{equation}
Remaining terms of the action can be arranged in a matrix form as
 $\Phi^T M^2\Phi/2 $, where
\begin{align}
 \Phi=\left(
\begin{array}{c}
 \hat{\chi}  \\
 \phi _3 \\
 h
\end{array}
\right),
\end{align}
is a scalr field triplet and $M$ is the mass matrix of this scalar field, which may be decomposed into unperturbed and perturbed parts as,
\begin{align}
 M^2=M_0^2 +\Delta M^2 .
\end{align}
 The diagonalization of unperturbed 
 mass matrix, considering only leading order terms, gives three eigenvalues which can be identified with the mass of Goldstone type mode, Higgs particle
and graviton, with the eigen functions as $\tilde{\chi}$, $\tilde{\phi_3}$ and $\tilde{g}$, respectively. Now we can write the action in term of these particles by inverse transformation. Here we are interested only in 
the interaction terms for the Higgs and the Weyl meson. These arise from kinetic energy term of Higgs boson  and coupling of Higgs with 
gravity in the action. So we can write the interaction Lagrangian as \cite{quantumweyl},
\begin{equation} \label{inter}
\begin{split} 
\mathcal{L}_{int} =&\frac{3}{4}f\beta_1[\tilde{\phi} _3{}^2S^{\mu }_{;\mu }+2v f \tilde{\phi} _3S^{\mu }
S_{\mu }+f \tilde{\phi} _3{}^2S^\mu S_\mu] \\ & -fS^\mu \tilde{\phi_3} \partial{\tilde{\phi}_3} +
f^2 v\tilde{\phi_3} S_{\mu } S^\mu+\frac{1}{2}f^2\tilde{\phi_3}^2 S^\mu S_{\mu }.                  
\end{split}    
\end{equation}

So far the values of parameters $f$ and $\beta_1$ are unconstrained. 
For application to collider physics,
we assume that $f$ is sufficiently small and $\beta_1$ is sufficiently large
 such that $f\beta_1$ is of order unity. In this case the 
mass of Weyl meson is small and it may produce 
observable effects at colliders such as LHC. We only need to consider the first term in Eq.(\ref{inter}). 
It may be represented diagrammatically as, 
\begin{figure}[ht!]
\begin{center}
\scalebox{0.3}{\includegraphics*[angle=0,width=\textwidth,clip]{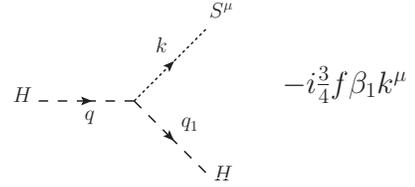}} 
\caption{Higgs-Weyl meson Interaction.}
\end{center}
\end{figure}
\subsection{Real Weyl meson production at LHC}
\begin{figure}[h!]
\begin{center}
\scalebox{0.3}{\includegraphics*[angle=0,width=\textwidth,clip]{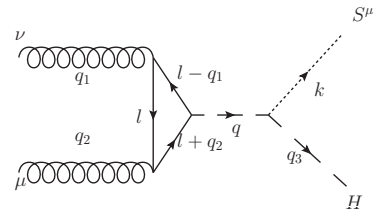}} 
\caption{Production of Higgs Boson and real Weyl meson at LHC.}
\label{weyl-higgs}
\end{center}
\end{figure}

The Weyl meson may be produced at 
LHC through its interaction with the Higgs 
particle. At LHC gluon-gluon fusion is the dominant process for the Higgs
 production \cite{Dawson,collider}. The corresponding Feynman diagram for
Weyl meson production 
 is shown in Fig.\ref{weyl-higgs}.

Let $\epsilon_\mu$ be the polarisation vector of the 
outgoing Weyl meson, then the matrix element for this process can be written as,
 \begin{equation*}
  \mathcal{M}\propto k^\mu\epsilon_\mu \\,
 \end{equation*}
hence,
\begin{equation*}
 \sum |\mathcal{M}|^2\propto\sum |k^\mu\epsilon_\mu|^2=0\\,
\end{equation*}
due to transversality condition \cite{chengli,collider} for real vector meson. Hence no real Weyl meson
 can be produced for this type of interaction at colliders.
\subsection{Virtual Weyl meson and associated Higgs boson production at LHC}
\begin{figure}[h!]
\begin{center}
\scalebox{0.55}{\includegraphics*[angle=0,width=\textwidth,clip]{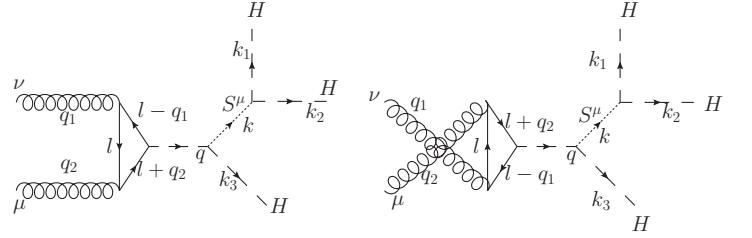}}
\caption{Virtual Weyl meson and three Higgs production at LHC.}
\label{thrihiggs}
\end{center}
\end{figure}

We have found that the amplitude for real Weyl meson production at LHC is zero. 
However the probability to produce a virtual Weyl meson may not be negligible.
This might allow us to indirectly infer its presence at LHC. Here we consider the Weyl meson 
produced as an intermediate virtual particle, which gives two real Higgs bosons as the final particles. This will lead to three Higgs production at LHC. 
We consider only the leading order diagrams of this process, shown in Fig.\ref{thrihiggs}
\\
The Higgs and the Weyl meson's propagators can be written as \cite{quantumweyl}
\begin{align}
\text{Higgs boson}      \,- - - - -  \quad \frac{-i}{q^2-m_H^2}\\
\text{Weyl Meson} \; ........  \quad \frac{-i}{k^2-M_S^2+i\epsilon} \left(g_{\mu\nu}-\frac{k_\mu k_\nu}{M_S^2}\right)
\end {align}
Here we write the Weyl meson's propagator in unitary gauge. Using these propagators the complete scattering matrix element for this process can be written as,
\begin{widetext}
\begin{equation}
\begin{split}
 \mathcal{M}  =&(-i g_s)^2 (\frac{-i y_t}{\sqrt{2}}) i^3 Tr[t_a t_b] \epsilon_\nu(\lambda_1,q_1) \, \epsilon_\mu (\lambda_2,q_2) 
              \int \frac{d^4l}{(2\pi)^4}\frac{1}{D_1 D_2 D_3} \, Tr\left[(\not{l}+\not{q_2}+m)\gamma^\mu(\not{l}+m)\gamma^\nu(\not{l}-\not{q_1}+m)\right.\\
              &\left.+(-\not{l}+\not{q_1}+m)\gamma^\nu(-\not{l}+m)\gamma^\mu(-\not{l}-\not{q_2}+m)\right]  \frac{-i}{q^2-m_H^2} 
              \left(-i\frac{3}{4}f \beta_1 k^\alpha \right) \,\frac{-i}{k^2-M_S^2+i\epsilon} \left(g_{\mu\nu}-\frac{k_\alpha k_\beta}{M_S^2}\right)\left(-i \frac{3}{4}f \beta_1 k^\beta \right) ,\\
\end{split}
\end{equation} 
\end{widetext}
where
\begin{equation*}
\begin{split}
D_1 & =l^2-m^2, \quad D_2 =(l-q_1)^2-m^2,\\
 D_3 & =(l+q_2)^2-m^2.
\end{split}
\end{equation*}
Here $m$ is the mass of quark in the loop and the most dominant contribution comes from the 
top quark. We follow the procedure given in Ref.\cite{bentvelsen} to do the loop integration over $l$ and get
\begin{equation}
\begin{split}
\mathcal{M}=\mathcal{M}_1\frac{-i}{q^2-m_H^2}&\left(-i\frac{3}{4}f \beta_1 k^\alpha \right) 
\,\frac{-i}{k^2-M_S^2}\\
& \left(g_{\mu\nu}-\frac{k_\alpha k_\beta}{M_S^2}\right)\left(-i \frac{3}{4}f \beta_1 k^\beta \right) ,\\
\end{split}
\end{equation}
where
\begin{equation*}
\begin{split}
 \mathcal{M}_1&=a(\epsilon_1.\epsilon_2-\frac{2}{\hat{s}}a(\epsilon_1.q_2)(\epsilon_2.q_1)\\
 a &=\frac{m}{2\pi^2}\left[1+(1-\lambda)f(\lambda)\right]\\
\end{split}
\end{equation*}
and $\epsilon_i=\epsilon_\mu(\lambda_i,k_i)$ are the polarisation vector of the gluons having spin
 state $\lambda_i$. We average over initial spin state of protons and colour states of gluons. Furthermore $g_s^2=8\pi\alpha_s$ and $\frac{y_t}{\sqrt{2}}=(\sqrt{2}G_fm^2)^{1/2}$. The colour trace 
term is $Tr[t_at_b]=\frac{1}{2}\delta_{ab}$. After putting all these factors and 
summing over spin state of gluons $\lambda_1,\lambda_2$, we get the squared amplitude for this process as, 
\begin{equation}
\sum_{\lambda_1,\lambda_2}|{\mathcal{M}}|^2=\frac{G_f \, \alpha_s^2 \, \hat{s}^2}{288 \sqrt{2} 
\pi^2}\left(\frac{3}{4}f\beta_1\right)^4 \, \frac{|A(\lambda)|^2 \left[s_{12}-\frac{s_{12}^2}{M_S^2}\right]^2}
 {(\hat{s}-m_H^2)^2 (s_{12}-M_S^2)^2},
\end{equation}
where Mandelstam variables are defined as,
\begin{align}
\hat{s}=(q_1+q_2)^2=q^2, \quad   s_{12} =(k_1+k_2)^2 \,=k^2
\end{align}
and
\begin{equation}
\begin{split}
 \lambda  & =\frac{4 m^2}{\hat{s}}, \quad \beta  =\sqrt{1-\lambda}\\
A(\lambda)& =\frac{3}{2} \lambda [1+(1-\lambda)f(\lambda)]\\
f(\lambda)& = \left\{ \begin{array}{l c}
arcsin^2(\frac{1}{\sqrt{\lambda}})   \qquad  \lambda \geq 1\\
-\frac{1}{4} \left[ln(\frac{1+\beta}{1-\beta})-i \pi \right] \qquad \lambda<1.
\end{array} \right.
\end{split}
\end{equation}
\subsection{Cross-Section}
The cross-section for this process at parton level can be written in 
CM frame as \cite{halzen},
\begin{equation}
\begin{split}
  \hat{\sigma} & =\frac{1}{2\hat{s}} \int  \sum|\mathcal{M}|^2 d\Phi_3 ,
\end{split}
\end{equation}
where $d\Phi_3$ is the three-body phase space.

Three body phase space integration can be done easily by decomposing it into two body 
phase space integral (see e.g \cite{phase,hitoshi}). 

\begin{figure}[ht]
\begin{center}
\scalebox{0.5}{\includegraphics*[angle=0,width=\textwidth,clip]{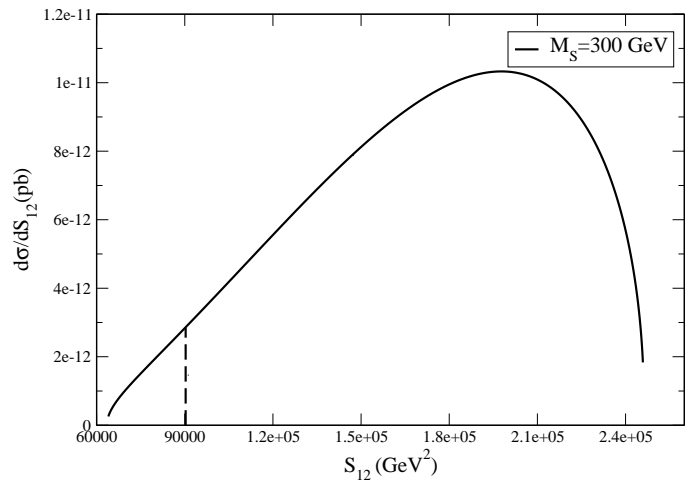}} 
\caption{Differential cross-section of three Higgs production at 7 TeV of CM energy for $M_S=300 GeV$. The dashed line in this figure indicates the position where
$S_{12}=M_S^2 $.}
\label{diffcs}
\end{center}
\end{figure}

\begin{figure}[ht]
\begin{center}
\scalebox{0.5}{\includegraphics*[angle=0,width=\textwidth,clip]{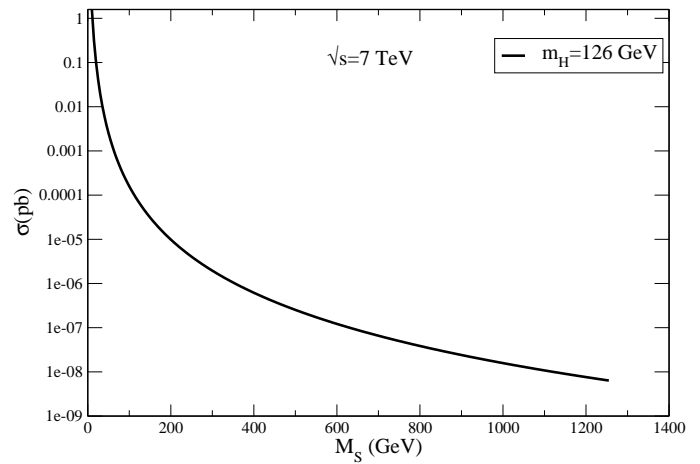}} 
\caption{Leading order three Higgs production cross-section at 7 TeV of CM energy for $m_H=126 GeV$.}
\label{3higgs}
\end{center}
\end{figure}

Total cross-section for this process is given by
\begin{equation}
 \sigma=\int  dx_1 dx_2 f(x_1) f(x_2) \hat{\sigma} ,\\
\end{equation}
where $x_1$ and $x_2$ are momentum fraction of gluons and 
$\hat{s}(= s x_1 x_2 $) is the square of total CM energy of the system at parton level. 
Taking $x_1x_2=T$ and $x_2$ as the independent variables \cite{collider}, the total cross-section become
\begin{equation}
\sigma=\int_{\frac{9 m_H^2}{s}}^1 dT \int_T^1  \, \frac{dx_2}{x_2}f(\frac{T}{x_2},Q) \, f(x_2,Q) \, \hat{\sigma}\\
\label{cs}
\end{equation}
Here $f(x,Q)$ is the parton distribution function(PDF) for the gluons.
 We also find the differential cross-section for this process to see the behaviour at on mass-shell
condition of Weyl meson,
\begin{align}
 \frac{d\sigma}{ds_{12}}= \int_{\frac{9 m_H^2}{s}}^1 dT \int_T^1  \, \frac{dx_2}{x_2} f(\frac{T}{x_2},Q) \, f(x_2,Q) \,\frac{d\hat{\sigma}}{ds_{12}}.
\label{dcs}
\end{align}
The differential cross-section for the three Higgs production, $d\sigma/dS_{12}$, is shown in Fig.\ref{diffcs}. The total cross-section, $\sigma$, is shown in Fig.\ref{3higgs}.
In these calculations we have used the CTEQ PDFs\footnote{http://www.phys.psu.edu/~cteq/}. 

We calculated the three Higgs production cross-section by setting $f\beta_1=1$, which is the maximum allowed value of the coupling within the 
perturbation region. It is clear from Fig.\ref{diffcs} that the differential cross-section varying smoothly with $S_{12}$ even in the vicinity of $S_{12}=M_S^2 $. 
This is because the production cross-section for real Weyl meson is negligible in this theory.

We find in Fig.\ref{3higgs} that the total cross-section for three Higgs production varies from $1pb$ at $M_S\approx10\,GeV$ to $10^{-8}pb$ at $M_S\approx1200\,GeV$.
 Hence the cross-section falls
sharply with increase in the mass of the Weyl meson. For comparison, the production cross-section for single Higgs boson with $m_H=126$, is roughly 
$3pb$. Hence for small values of the Weyl meson mass, the three Higgs production cross-section is not negligible compared to single Higgs cross-section. The background to 
the three Higgs cross-section would arise from the top quark loop. This will include two additional vertices of top quark with the Higgs boson. Hence the process is suppressed by 
$(\alpha_{Yt})^2\approx 10^{-2}$, compared to single Higgs production. We, therefore, conclude that for small $M_S<100\, GeV$, it may be possible to extract the contribution due to
Weyl meson to three Higgs production cross-section at LHC.
\section{Weyl meson as a Dark Matter Candidate}
The Weyl meson acts as a dark matter candidate since it interacts only with Higgs boson and no other SM particles. In Sec. II we wrote down the various interaction terms for
 Weyl meson. At very early time in the Universe, when temperature was very high, Weyl Meson may remain in 
thermal equillibrium with other particles by these interactions.
Its number density keeps on changing by the coannihilation
 process. But as the temperature of the Universe drops down to certain level, due to expansion, its interaction rate become less than the expansion rate and 
coannihilation process ceases. It gets decoupled from other particles and its number density
 freezes out at this temperature and remains as relic density. In this section we calculate the coannihilation rate of the Weyl meson  and 
find the possible values of parameters $\beta_1$ and $M_S$  by constraining its abundance equal to the observed dark matter abundance.
\subsection{Coannihilation rate of the Weyl meson}
\begin{figure}[h]
\begin{center}
\scalebox{0.5}{\includegraphics*[angle=0,width=\textwidth,clip]{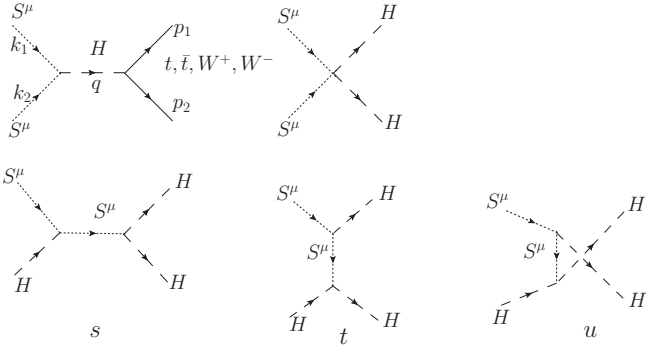}} 
\caption{Weyl meson annihilation processes.}
\label{annihilation}
\end{center}
\end{figure}

We consider only the dominant processes which will contribute to the total coannihilation rate of the Weyl meson, as shown in Fig.\ref{annihilation}.
\\
 Interaction rate for annihilation process can be written as
\begin{center}
           $\qquad \Gamma=n_{eq}<\sigma v>$,
\end{center}
where $<\sigma v>$ is the thermally averaged annihilation cross section given by, \cite{gondolo,gondolo1}
 \begin{equation}
<\sigma v>=\frac{g_s^2 T }{32 n_{eq}^2\pi^4} \int_{4m_s^2}^{\infty} ds\,\sigma \left(P_{12}^2\,
 \sqrt{s}\right)\, K_1\left(\frac{\sqrt{s}}{T}\right).
 \end{equation}
Here $K_1(x)$ is the modified Bessel function of second kind of order one, $T$ is the temperature at which proces is occuring, 
$g_s$ is the number degrees of freedom
of the Weyl meson and $P_{12}$ is the four momentum in center of mass frame of incoming particles, which is given by
\begin{equation}
P_{ij}=\frac{\sqrt{s}}{2}\sqrt{1-2\left(\frac{m_i^2+m_j^2}{s}\right)+\frac{(m_i^2-m_j^2)^2}{s^2}}.
\end{equation}

Let $\epsilon$ be the polarisation vector of incoming Weyl meson, then the scattering amplitude for annihilation of two Weyl meson into fermion
pair can be written as
\begin{align}
 \mathcal{M}_1=& \nonumber \frac{3}{2}\,f^2 \,\beta_1(\sqrt{2}G_f)^{\frac{1}{2}} m_t \epsilon_\mu \epsilon_\mu\\
&*\left(\frac{-i}{q^2-m_H^2+im_H\varGamma_t}\right)\bar{u}(p_1)v(p_2)\\
\sum_{spin}|{\mathcal{M}_1}|^2=& \nonumber \frac{1}{4}\left(\frac{(f^2\beta_1v)^2\sqrt{2} G_f m_t^2}{(s-m_H^2)^2+m_H^2
\varGamma_t^2}\right) \left(\frac{s^3}{2M_S^4} \right) \\
&*(1+12\lambda_s^2-4\lambda_s) (1-4\lambda_b),
\end{align}
where $\lambda$ is defined as,
\begin{center}
 $\lambda_i=\frac{M_i^2}{s}$.
\end{center}
The top quarks, being massive, will have the dominant contribution in this case. 
Now we can write the cross section for this process in CM frame of incoming particles,
\begin{align}
\sigma=&\frac{\sum|{\mathcal{M}_1}|^2 \sqrt{1-4\lambda_b}}{32\pi P_{12}\sqrt{s}},\;P_{12}=\frac{\sqrt{s}}{2}\sqrt{1-\frac{4 M_S^2}{s}}.
\end{align}
Hence the resulting coannihilation rate for this process become
\begin{align}
 \Gamma_1=& \nonumber \frac{g_s^2(f^2\beta_1v)^2 G_f m_t^2T}{8192\sqrt{2}\pi^5 M_S^4 n_{eq}} \\
&\int_{4M_S^2}^{\infty} 
ds\frac{s^{\frac{7}{2}}\sqrt{1-4\lambda_s} (1-4\lambda_s+12\lambda_s^2)}{(s-m_H^2)^2+m_H^2\varGamma_t^2}
 K_1(\frac{\sqrt{s}}{2}).
\end{align}
For numerical integration, we change the variable $s$ to $x$ as, 
\begin{align}
 \frac{s}{4T^2}-\frac{M_S^2}{T^2}&=x^2\\
 \frac{M_S}{T}=y
\end{align}
and write the expression for $n_{eq}$ in term of modified Bessel function of second kind of order two ($K_2(x)$) (see \cite{drees}) 
\begin{align}
 n_{eq}=\frac{g_s}{2\pi^2} T M_S^2 K_2(y).
\end{align}
We can simplify the expression for coannihilation rate in the limit $s>>m_H^2$,
as we are considering these process very early in the Universe, when temperature was very high.
 Considering only those terms which are linear in $\lambda_i$ results in coannihilation rate
\begin{align}
 \Gamma_1=\frac{g_s(f^2\beta_1v)^2 G_f m_b^2T}{64\sqrt{2}\pi^3 M_S K_2(y) y^5} \int_0^{\infty}
 dx\, x^4 K_1\left(2\sqrt{x^2+y^2}\right). 
\end{align}
Similarly we can write the coannihilation rate for second process, where Weyl meson annihilate into the Higgs pair, as
\begin{align}
 \Gamma_2=\frac{g_sM_S \left(f^2\beta_1\right)^2}{256 \pi^3 K_2(y)y^7} \int_0^{\infty} dx\,x^4
 \left(x^2+y^2\right)^2 K_1\left(2\sqrt{x^2+y^2}\right).
\end{align}

Next process is the annihilation of Weyl meson with Higgs particle, for which we have $s,t$ and $u$ channel processes.
 Matrix element of these processes will be sum of all these channels,  
 i.e $\mathcal{M}=\mathcal{M}_s+\mathcal{M}_t+\mathcal{M}_u$\\
where
\begin{align}
\mathcal{M}_s&=\nonumber \frac{9}{8}f^3\beta_1^2 v\left(\frac{\epsilon_{\mu}}{(s-M_S^2)}(g^{\mu\nu}-\frac{k^{\mu}k^{\nu}}{M_S^2})k_{\nu}\right)\\
             &=\nonumber\frac{9}{8}f^3\beta_1^2 v\left(\frac{(\epsilon.k)(1-\frac{s}{M_S^2})}{(s-M_S^2)}\right)\\
             &=-\frac{9}{8}f^3\beta_1^2 v \frac{\epsilon.k_s}{M_S^2}.
\end{align}
Similarly for other channels
\begin{equation}
 \mathcal{M}_t=-\frac{9}{8}f^3\beta_1^2 v \frac{\epsilon.k_t}{M_S^2},\;\; \mathcal{M}_u= -\frac{9}{8}f^3\beta_1^2 v \frac{\epsilon.k_u}{M_S^2}.
\end{equation}
We used the subscript on $k$ to define it in different channels.

The amplitude of this process contains square term of each channel's matrix element and also the cross terms, 
which we can write by using 
\begin{align*}
(\epsilon.k_s)(\epsilon.k_t)=\frac{(p.k_s)(p.k_t)}{M_S^2},
\end{align*}
and \begin{align}
     (\epsilon.k_s)^2&=\nonumber k_s^{\mu}(-g_{\mu\nu}+\frac{p_{\mu}p_{\nu}}{M_S^2})k_s^{\nu}\; =-k_s^2+\frac{(p.k_s)^2}{M_S^2},\\
                  p.k_s&=\frac{k_s^2+p^2-q^2}{2}=\frac{s+M_S^2-m_H^2}{2}.  
\end{align}
We obtain similar expressions for other channels by replacing $s$ with $t$ and $u$, accordingly.

The cross section for this process can be written as, 
   \begin{align}
    \sigma&=\nonumber \frac{1}{4P_{12}\sqrt{s}} \int \sum_{spin}|\mathcal{M}|^2 d\Phi\\
          &=\frac{1}{4P_{12}\sqrt{s}} \int \sum_{spin}(|\mathcal{M}_s|^2+|\mathcal{M}_t|^2+|\mathcal{M}_u|^2\\
 &\nonumber +2\mathcal{M}_s \mathcal{M}_t
         +2\mathcal{M}_s \mathcal{M}_u+2\mathcal{M}_t \mathcal{M}_u) d\Phi.
   \end{align}     

Following the same procedure as before, we define the new variables $x,\,y$ and $y_1$ such that
 \begin{align}
y_1&=\nonumber\frac{M_S+m_H}{T}, \quad y=\frac{M_S}{T}\\
  x^2&=\frac{s-(M_S+m_H)^2}{T^2},
 \end{align}
and write the interaction rate $\Gamma_3$ in term of these variables. Here again we simplify the 
expression in the limit $s>>m_H^2$, and consider only linear terms in $M_S/s$. So $\Gamma_3$ will be of this form
\begin{align}
\Gamma_3&=\frac{g_s^2}{32 n_{eq}^H\pi^4} \int_0^{\infty} dx\,F(x,y,y_1) K_1(\sqrt{x^2+y_1^2})
 \end{align}
where we have absorbed all other quantities in the function $F(x,y,y_1)$.
Total annihilation rate $\Gamma$ of Weyl meson is the sum of all three interaction rates, i.e
\begin{align}
 \Gamma=\Gamma_1+\Gamma_2+\Gamma_3
\label{gama}
\end{align}

 \subsection{Decoupling of Weyl meson and its abundance}
As the Universe expands, temperature drops down and probability of annihilation of Weyl meson gets decreased. 
At some temperature annihilation rate become equal to the expansion rate of Universe, known as freeze out temperature $T_f$.
 Below this temperature there is no annihilation process, Weyl meson decouples from the surrounding and moves freely in the Universe.
 At freeze out temperature, the number density of Weyl meson $n_f$ will get frozen and decrease only because of 
expansion of the Universe.

At freeze out temperature $\quad\Gamma=\mathcal{H}$,\\
i.e 
\begin{align}
 n_f\langle\sigma v\rangle_f=\frac{1.66\sqrt{g_*}T_f^2}{m_{pl}},
\end{align}
where 
\begin{align}
 \langle\sigma v\rangle_f=\left(\frac{\Gamma}{n_{eq}}\right)_{T_f},
\label{sgmav}
\end{align}
is the thermally averaged 
annihilation cross-section at freeze out temperature $T_f$ and $g_*$ is the number of relativistic degrees of freedom
of the all species present at that time.
 We can write it in invariant form as \cite{kolb}
\begin{align}
\left(\frac{n}{S}\right)_f=\frac{74.7\sqrt{g_*}}{2\pi^2 m_{pl}\langle\sigma v\rangle_f g_* T_f},
\end{align}
where we have used the formula for entropy density $S$ of Universe given by,
\begin{align}
 S=\frac{2\pi^2}{45} g_*T^3.
\end{align}
As this quantity remain constant, we obtain,
\begin{align}
\left(\frac{n}{S}\right)_{today}=\left(\frac{n}{S}\right)_{freeze}
\end{align}
Hence we can write the abundance of Weyl meson in present Universe,
\begin{align}
 \Omega_s=\frac{74.7 S_0 M_S}{2\pi^2 m_{pl}\sqrt{g_*}T_f \rho_c \langle\sigma v\rangle_f}
\label{omega}
\end{align}
where $S_0$ is the present value of entropy density, $\rho_c$ is the critical density of the Universe and value of $g_*\approx106.75$
for processes at very high temperature.

Using $<\sigma v>_f$ from Eq.\eqref{sgmav} and $f=M_S/m_{pl}$, we solve Eq.\eqref{omega} numerically for different value of $\beta_1$
and $M_S$. We consider the values of $\Omega_s$ to be equal to present dark matter density($\Omega_s\approx0.3$).
 In all calculations we take the mass of Higgs boson $m_H=126GeV$, as claimed by CMS and ATLAS groups. 
\begin{figure}[h]
\begin{center}
\scalebox{0.32}{\includegraphics*[angle=0,width=1.5\textwidth,clip]{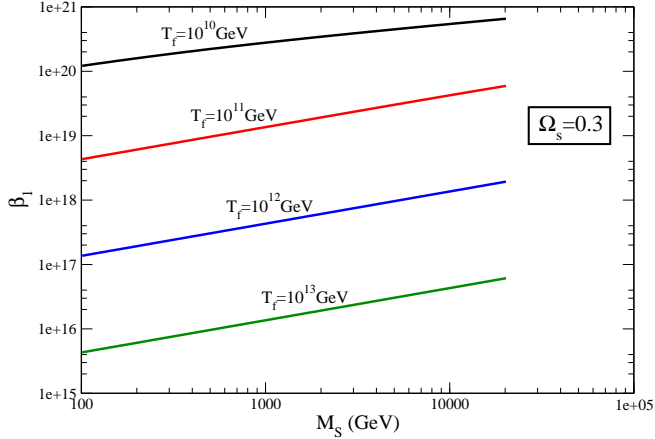}} 
\caption{Coupling parameter $\beta_1$, as obtained by constraining the relic density of Weyl meson to be equal to 
the dark matter density($\Omega_s=0.3$) for different freeze out temperature $T_f$ and mass $M_S$.}
\label{bms}
\end{center}
\end{figure}

The values of the parameter $\beta_1$ and $M_S$ for decoupling temperature $T_f=10^{10}$ to $10^{13}\,GeV$ are shown in Fig.\ref{bms}. We find, as expected, the lower
 $\beta_1$ leads to higher decoupling temperature. Hence for small value of coupling parameter Weyl meson get decoupled very early in the Universe. 
Furthermore $\beta_1$ also increase with $M_S$. 
\section{Decay Of Weyl Meson}
In the previous section we have constrained the values of coupling parameters from the coannihilation rate and abundance of Weyl meson. 
But Weyl meson may also have the probability to decay into Higgs particles. So it may be possible that after the decoupling all the Weyl meson 
particles have already been decayed and today they may not be present in the Universe. In this section we calculate the 
decay rate of the Weyl meson and constraint its life time to be equal to the age of Universe. We show that the certain range of parameters
 $\beta_1$ and $M_S$ satisfiies these conditions.

\begin{figure}[h]
\begin{center}
\scalebox{0.2}{\includegraphics*[angle=0,width=1.5\textwidth,clip]{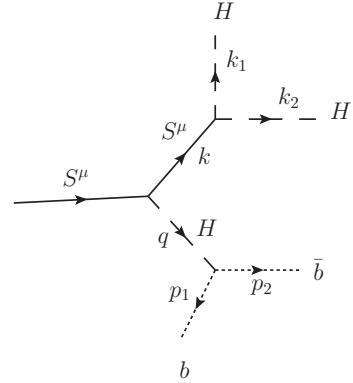}} 
\caption{Decay process of Weyl meson.}
\label{wdecy}
\end{center}
\end{figure}

 We consider the simplest process for the
decay of Weyl meson into pairs of Higgs particles and fermions, as shown in Fig.\ref{wdecy}

The amplitude for this process is given by
\begin{align}
\mathcal{M}=&\nonumber \frac{9}{8} f^3\beta_1^2 \mathbf{v}(\sqrt{2} G_f)^\frac{1}{2} m_b 
\frac{\epsilon^\mu k^\nu}{(k^2-m_s^2)(q^2-m_H^2)} \\&*\left(g_{\mu\nu}-\frac{k_\mu k_\nu}
{m_s^2}\right) \bar{u}(p_1) v(p_2)\\
\sum_{spin}|{\mathcal{M}}|^2= &\frac{A^2(f^3\beta_1^2)^2(\epsilon.k)^2 (1-k^2)^2 (4 p_1.p_2-4m_b^2)}{(k^2-m_s^2)^2 (q^2-m_H^2)^2},
\end{align}
where $A$ contains all the constant. Now we define four momenta square  $s_{12},s_{34}$, such that
\begin{align}
k^2=(q_1+q_2)^2=s_{12},\quad q^2=(p_1+p_2)^2=s_{34}.
\end{align} 
In term of these variables, we find
\begin{align}
p_1.p_2=(\frac{s_{34}}{2}-m_b^2),\quad (\epsilon.k)^2=-s_{12}+\frac{m_s^2+s_{12}-s_{34}}{m_s^2}
\end{align}
and hence the amplitude of this process becomes function of the variables $s_{12}$ and $s_{34}$.\\
The decay rate for this process is
\begin{align}
\Gamma=\frac{1}{2 M_S}\int \sum_{spin}|{\mathcal{M}}|^2 d\Phi,
\label{drate}
\end{align}
where $d\Phi$ is the four dimensional phase space integral and again can be decomposed into product of
two dimensional phase space integral\cite{hitoshi}
\begin{align*}
\int d\Phi=\int \frac{ds_{12}}{2\pi}\frac{ds_{34}}{2\pi} d\Phi_2(\hat{p_1},\hat{p_2})d\Phi_2(\hat{q_1},\hat{q_2}) d\Phi_2(\hat{q},\hat{k})
\end{align*}
Here $d\Phi_2(\hat{p}_i,\hat{p}_j)$ is the two dimensional phase space integral in rest frame of 
particles having momentum $\hat{p}_i$ and $\hat{p}_j$.

  If Weyl meson is a suitable dark matter candidate then it must be present in today's Universe. 
So its life time($\tau=1/{\Gamma}$) must be at least equal to or greater than the age of the Universe. We determine $\Gamma$ numerically from Eq.(\ref{drate}) and find those values
of $\beta_1$ and $M_S$ for which the life time of Weyl meson is equal to the age of the Universe $(\tau\approx 10^{42}GeV^{-1})$.

We have considered only the simplest process of the Weyl meson's decay, as we are interested only in its decay rate to constrain the value of parameter $\beta_1$ and mass $M_S$. 
Final decay products can also be photons or electron-positron pairs. There are many observations showing the excess of positrons in cosmic rays \cite{pamela,fermi}.
This excess of positrons may also be due to the decay products of the Weyl meson. We postpone the detailed analysis and fitting of the data for future research.

\begin{figure}[ht]
 \scalebox{0.5}{\includegraphics*[angle=0,width=\textwidth,clip]{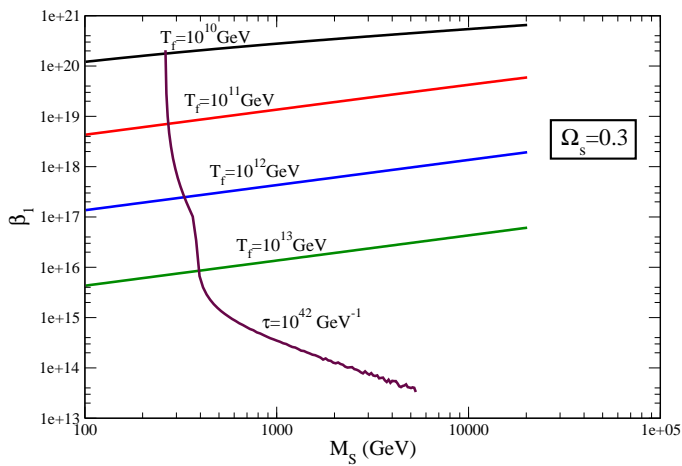}}
\caption{Possible values of parameter $\beta_1$ and mass $M_S$ of Weyl meson to solve the dark matter problem at given decoupling temperature $T_f$.}
\label{relic}
\end{figure}

Now we have the posible range of parameter $\beta_1$ and mass $M_S$ from its relic density condition and its decay rate constraint. If Weyl meson is the dark matter candidate then it must
 satisfy both of these conditions. Combining these two results we have obtained the possible values of $\beta_1$ and $M_S$ for the Weyl meson, as shown in Fig.\ref{relic}. 

Here decoupling temperature $T_f$ is still an unknown parameter, but the intersting result which we have obtained is that the
mass of the Weyl meson can not be greater than $500\,GeV$ and the parameter $\beta_1$ has minimum value $\approx10^{15}$. These are the
values which the Weyl meson can have if it is a very weakly interacting particle, such that it decouple immediately 
after the reheating of the Universe($T_{rh}\approx10^{14}\,GeV$). If $\beta_1$ is smaller than $10^{15}$ then the Weyl meson decouples
before the reheating of Universe and no constraint can be imposed on its relic density. We also find that the low mass range of the Weyl meson,
which may be observable in colliders, is allowed by cosmological constraints.

\section{conclusion}
Generalized locally scale invariant Standard Model leads to the direct coupling between the Weyl meson and the Higgs boson\cite{quantumweyl}. Weyl meson has been introduced in 1929 \cite{weyl}. 
In this paper we have discussed the implications of the Weyl meson in collider physics and in cosmology in detail.
 We find that in the colliders real Weyl meson can not be produced for such type of interaction, but it can be produced virtually and leads to the production of three Higgs particles.
 We have discussed the three Higgs production process at LHC which can give us the signature of the Weyl meson for some finite value of coupling parameter. Furthermore, in this model 
Weyl meson acts as a dark matter candidate. We have discussed how for certain range of coupling parameter $\beta_1$ and mass $M_S$ Weyl meson can solve the dark matter problem.
Although its decoupling temperature $T_f$ is still an unknown parameter, but we have found the upper bound for the Weyl meson mass $M_S$ and lower bound for its parameter $\beta_1$
 ( $M_S<500\,GeV\;, \beta_1>10^{15})$ so that it is in thermal equillibrium with the cosmic fluid at some early time after the reheating of the Universe.
In future we hope to have some data from colliders and from cosmic observations to test this model and to fix the values of these parameters.
\section*{Acknowledgments} 
I thank Prof. Pankaj Jain for many useful discussions and valuable comments. I also thank Nishtha Sachdeva for collaboration during the early stages
of this work. I sincerely acknowledge CSIR, New Delhi for financial assistance.

\end {document}